\documentclass[10pt,twocolumn,showpacs,amsmath,amssymb,floatfix,superscriptaddress]{revtex4}
\usepackage{graphicx}
\usepackage{float}
\usepackage{dcolumn}
\usepackage{bm}
\usepackage{color}
\usepackage{txfonts}
\usepackage{microtype}
\usepackage{mathrsfs}
\usepackage{amsmath}
\usepackage{extarrows}
\usepackage{mathtools}
\usepackage{xcolor}
\usepackage{slashed}
\usepackage{amsmath}

\begin{document}

	\title{Vortex information in multiphoton scalar pair production}

	\author{Hong-Hao Fan}
	\affiliation{Key Laboratory of Beam Technology of the Ministry of Education, and School of Physics and Astronomy, Beijing Normal University, Beijing 100875, China}

  \author{Cui-Wen Zhang}
	\affiliation{Key Laboratory of Beam Technology of the Ministry of Education, and School of Physics and Astronomy, Beijing Normal University, Beijing 100875, China}

	\author{Suo Tang}\email{tangsuo@ouc.edu.cn}
	\affiliation{College of Physics and Optoelectronic Engineering, Ocean University of China, Qingdao, Shandong, 266100, China}

	\author{Bai-Song Xie}\email{bsxie@bnu.edu.cn}
	\affiliation{Key Laboratory of Beam Technology of the Ministry of Education, and School of Physics and Astronomy, Beijing Normal University, Beijing 100875, China}
	\affiliation{Institute of Radiation Technology, Beijing Academy of Science and Technology, Beijing 100875, China}

\date{\today}
\begin{abstract}
Vortex information of scalar pair production in circularly polarized field is investigated in the multiphoton regime. We find that vortex orientation is related to the intrinsic orbital angular momentum of created particles associating with the helicity of absorbed photons, while the magnitude of the orbital angular momentum, i.e., the topology charge is determined by the number of absorbed photons. Moreover, the properties of particle creation and vortices formation can be understood by analyzing the pair production process in quasiparticle representation. This study provides new insights into the angular momentum transfer from field to particle in the scalar pair production process. It is expected that there are similar findings about vortex features for different spin alignment in electron-positron pair production in strong fields via the topology charge as a new freedom.
\end{abstract}
\maketitle

\section{Introduction}

The phenomenon of electron-positron pair production in vacuum~\cite{Heisenberg:1936nmg,Sauter:1931zz,Schwinger:1951nm,PhysRevD.44.1825,Fradkin:1991,Xie:2017xoj,DiPiazza:2011tq,Ruf:2008ahs} is closely analogous to ionization in atomic physics~\cite{Keldysh:1965ojf,Ma:2013,Troup:1972xkz,NBDelone:1994,RKopold:2002}.
Both processes are including above-threshold ionization, harmonic generation in perturbative multiphoton process~\cite{BECKER200235,Michael:2005,Muller:2009zzf,Aleksandrov:2021ylw} and the non-perturbative tunneling process~\cite{Schwinger:1951nm,Xie:2017xoj,Klaiber:023,Klaiber:2024}.
Besides, pairs production and ionization driven by varying dynamic fields exhibit many similar effects in momentum space,
including a series of multiphoton peaks in single circularly polarized field~\cite{Kohlfurst:2018kxg,Blinne:2013via,Li_2015,Fillion-Gourdeau:2017uss,Geng2020VortexSI}, and spiral structures in two counter-rotating electric field with time delay~\cite{Ngoko:2015,Pengel:2017,Li:2017qwd,Li:2018hzi,Li:2019rex,Majczak:2022xlv,Hu:2023pmz,Yusoff:2024}.

Wavefield phase front singularities, vortices, and associated non-integral phases have been introduced and studied in many  academic papers~\cite{PMDirac:1931,Aharonov,berry,Allen:1992zz,Beth:1936zz,Bialynicki-Birula:2016unl,Hebenstreit:2016xhn}. In the study of the photodetachment, phase vortices naturally appear in the electronic eigenstates of atoms, resulting in vortex structures in momentum space~\cite{BialynickiBirula1992TheoryOQ}. Recent research has found that vortex structures also appear in pairs production via the Sauter-Schwinger process~\cite{Bechler:2023kjx,Majczak:2024hmt}.

Moreover, in the perturbative multiphoton regime, the created particles exhibit distinct angular momentum features that are inherited from the photon helicity~\cite{Kohlfurst:2018kxg,Li_2015,Fan:2024vqi}. This suggests that the angular momentum of the created particles possesses intrinsic properties that may result in singularities in the amplitude phase, e.g., the vortices of the particle carry a quantized orbital angular momentum~\cite{Bialynicki-Birula:2016unl,Lloyd:2017,Bliokh:2017uvr}. This motivates our interest in studying vortices in pairs production in multiphoton-dominated processes under circularly polarized fields. For simplicity, we will consider relativistic scalar particle without spin. This is because that the wave packets with orbital angular momentum (OAM) exhibit fundamental topological and dynamical characteristics that have traditionally been associated primarily with spin~\cite{Bliokh:2007ec}.

In this work, the particle distribution function in momentum space is obtained by solving the Klein-Gordon equation and can be further reduced to a two-level system. 
We find that the vortex structures and the topological charge magnitude are determined by the absorbed photons.
Additionally, the evolution of vortices is investigated in the quasiparticle representation, which shows that momentum spectra not only exhibit the spiral structure but also the phase distribution shows the transition from the spiral pattern to the multi-vortex structures.

It should be noted that the natural units are employed throughout this paper with $\hbar = c = 1$.

This paper is organized as follows.
In Sec.~\ref{Sec:II}, the external field and particle distribution function are introduced, while the description of the vortex is reviewed.
In Sec.~\ref{Sec:Iv}, the numerical results are presented.
Finally, a brief summary is given in Sec.~\ref{summary}.

\section{Theoretical model}\label{Sec:II}

We adopt the following idealized model of circularly polarized electric field:
\begin{align}\label{Eq:1}
	\boldsymbol{E}\left(t\right) = E\sin^4\left({\omega t}/{2N}\right) \left[ \cos \left(\omega t\right) \boldsymbol{e_x} + \delta \sin\left(\omega t\right) \boldsymbol{e_y} \right],
\end{align}
where $E$ gives the external field amplitude in units of the critical field $E_\text{cr} = m^2/e$, $\omega$ is the frequency, $N$ is the number of cycles, and $ 0\leq t \leq 2N\pi/\omega$.
The polarization vector  $\boldsymbol{e_x} $ and $\boldsymbol{e_y} $ are along the $x$ and $y$ directions, respectively. The laser pulse propagates along the $\boldsymbol{e_z}$ axis, with $\lvert \delta \rvert$ representing the ellipticity, constrained by $-1 \leq \delta \leq 1$. For linear polarization, $\delta = 0$, while circular polarization corresponds to $\lvert \delta \rvert = 1$. The sign of $\delta$ determines the helicity, where $\delta = +1$ indicates right-handed circular polarization, and $\delta = -1$ indicates left-handed circular polarization.

When we neglect the backreaction, we can solve the Klein-Gordon equation for the scalar pairs production in the background electric field $\boldsymbol{E}\left(t\right)$ in Eq.~\eqref{Eq:1},
along with the associated gauge potential $\boldsymbol{A}\left(t\right)$, and $\boldsymbol E\left(t\right) = -\dot{\boldsymbol A}\left(t\right)$.
A natural formalism for this analysis is provided by Bogoliubov transformations~\cite{Kluger:1998bm}.

Given the initial creation and annihilation operators $a_{\boldsymbol{p}}, b_{\boldsymbol{p}}^{\dagger}$, the Bogoliubov transformation coefficients $\alpha_{\boldsymbol{p}}\left(t\right), \beta_{\boldsymbol{p}}\left(t\right)$ relate these operators to their time-dependent forms $\tilde a_{\boldsymbol{p}}\left(t\right),\tilde b_{-{\boldsymbol{p}}}^{\dagger}\left(t\right)$~\cite{Dumlu:2011rr,Akkermans:2011yn,Dunne:2022zlx}
\begin{align}
	\begin{bmatrix}
		\tilde a_{\boldsymbol{p}}\left(t\right) \\
		\tilde b_{-\boldsymbol{p}}^{\dagger}\left(t\right)
	\end{bmatrix}
	=
	\begin{bmatrix}
		\alpha_{\boldsymbol{p}}\left(t\right) &  \beta_{\boldsymbol{p}}^{*}\left(t\right)\\
		\beta_{\boldsymbol{p}}\left(t\right) &  \alpha_{\boldsymbol{p}}^{*}\left(t\right)
	\end{bmatrix}
	\begin{bmatrix}
		a_{\boldsymbol{p}} \\
		b_{-\boldsymbol{p}}^{\dagger}
	\end{bmatrix} \ ,
\end{align}
where $\left| \beta_{\boldsymbol{p}}\left(t\right)  \right|^2$ is the density of the created particles with canonical momentum $\boldsymbol{p}$.
The bosonic statistics require  $\left|\alpha_{\boldsymbol{p}}\left(t\right)\right|^2 - \left| \beta_{\boldsymbol{p}}\left(t\right)  \right|^2=1$.  The time-dependent Bogoliubov
coefficients satisfy the following relations~\cite{Li:2019rex}:
\begin{align}
	\frac{d\alpha_{\boldsymbol{p}}\left(t\right)}{dt} = -\Omega_{\boldsymbol{p}}\left(t\right) \beta_{\boldsymbol{p}}\left(t\right) e^{2i\int^t d \tau \omega_{\boldsymbol{p}}(\tau)}, \notag \\
	\frac{d\beta_{\boldsymbol{p}}\left(t\right)}{dt} =  -\Omega_{\boldsymbol{p}}\left(t\right) \alpha_{\boldsymbol{p}}\left(t\right) e^{-2i\int^t d \tau\omega_{\boldsymbol{p}}(\tau)},	
\end{align}
where  $\Omega_{\boldsymbol{p}}\left(t\right) = - {\dot{\omega}_{\boldsymbol{p}}\left(t\right)}/{2\omega_{\boldsymbol{p}}\left(t\right)}$. The time evolution of the system can be directly expressed  in terms of the Bogoliubov coefficients, whose dynamics are described by a two-level system. By setting $c_{\alpha, {\boldsymbol{p}}}=e^{-i\int^t d \tau \omega_{\boldsymbol{p}}(\tau)}\alpha_{\boldsymbol{p}}\left(t\right)$, $c_{\beta, {\boldsymbol{p}}} = e^{i\int^t d \tau \omega_{\boldsymbol{p}}(\tau)}\beta_{\boldsymbol{p}}\left(t\right) $, the evolution equation takes the form~\cite{Dunne:2022zlx}
\begin{align}
	i\frac{d}{d t}
	\begin{bmatrix}
		c_{\alpha, {\boldsymbol{p}}}\left(t\right) \\
		c_{\beta, {\boldsymbol{p}}}\left(t\right)
	\end{bmatrix} & =
	\begin{bmatrix}
		\omega_{\boldsymbol{p}}\left(t\right) & -i\Omega_{\boldsymbol{p}}\left(t\right) \\
		-i \Omega_{\boldsymbol{p}}\left(t\right) & -\omega_{\boldsymbol{p}}\left(t\right)
	\end{bmatrix}
	\begin{bmatrix}
		c_{\alpha, {\boldsymbol{p}}} \\
		c_{\beta, {\boldsymbol{p}}}
	\end{bmatrix}\label{eq:twolevelsys},
\end{align}
where $\omega_{\boldsymbol{p}}\left(t\right)=\sqrt{m^2 +\boldsymbol q^2\left(t\right)} = \sqrt{m^2 + \left[\boldsymbol{p} - e\boldsymbol{A}\left(t\right)\right]^2}$ denotes the single-particle energy
with kinetic momentum $\boldsymbol {q}\left(t\right)$ given by  $\boldsymbol{q}\left(t\right) = \boldsymbol{p} - e\boldsymbol{A}\left(t\right)$.
The single-particle distribution function in momentum space is expressed as
\begin{align}\label{distribution}
	f({\boldsymbol{p}}) = \lvert c_{\beta, {\boldsymbol{p}}} \rvert ^2.
\end{align}
It can be verified that the distribution function can be reduced to the well-known quantum Vlasov equation (QVE)~\cite{Blaschke:2017igl,Blaschke:2014fca}
\begin{align}\label{Eq:6}
	\frac{d}{d t} f\left(\boldsymbol{p}, t\right)  = & \frac{1}{2} \Omega_{\boldsymbol{p}}\left(t\right) \int_{-\infty}^{t} d t^{\prime} \Omega_{\boldsymbol{p}}\left( t^{\prime}\right)\Bigg\{\left[1+2 f\left(\boldsymbol{p}, t^{\prime}\right)\right]
	\times  \notag \\ & \cos \left[2 \Theta\left(\boldsymbol{p}, t^{\prime}, t\right)\right]\Bigg\},
\end{align}
where $ \Theta\left(\boldsymbol{p}, t^{\prime}, t\right) = \int_{t^{\prime}}^{t} d \tau \omega_{\boldsymbol{p}}\left( \tau \right) $.

\begin{figure*}[!t]
	\centering
	\setlength{\abovecaptionskip}{-0.cm}
	\setlength{\belowcaptionskip}{-0.cm}
	\centering\includegraphics[width=0.9\linewidth]{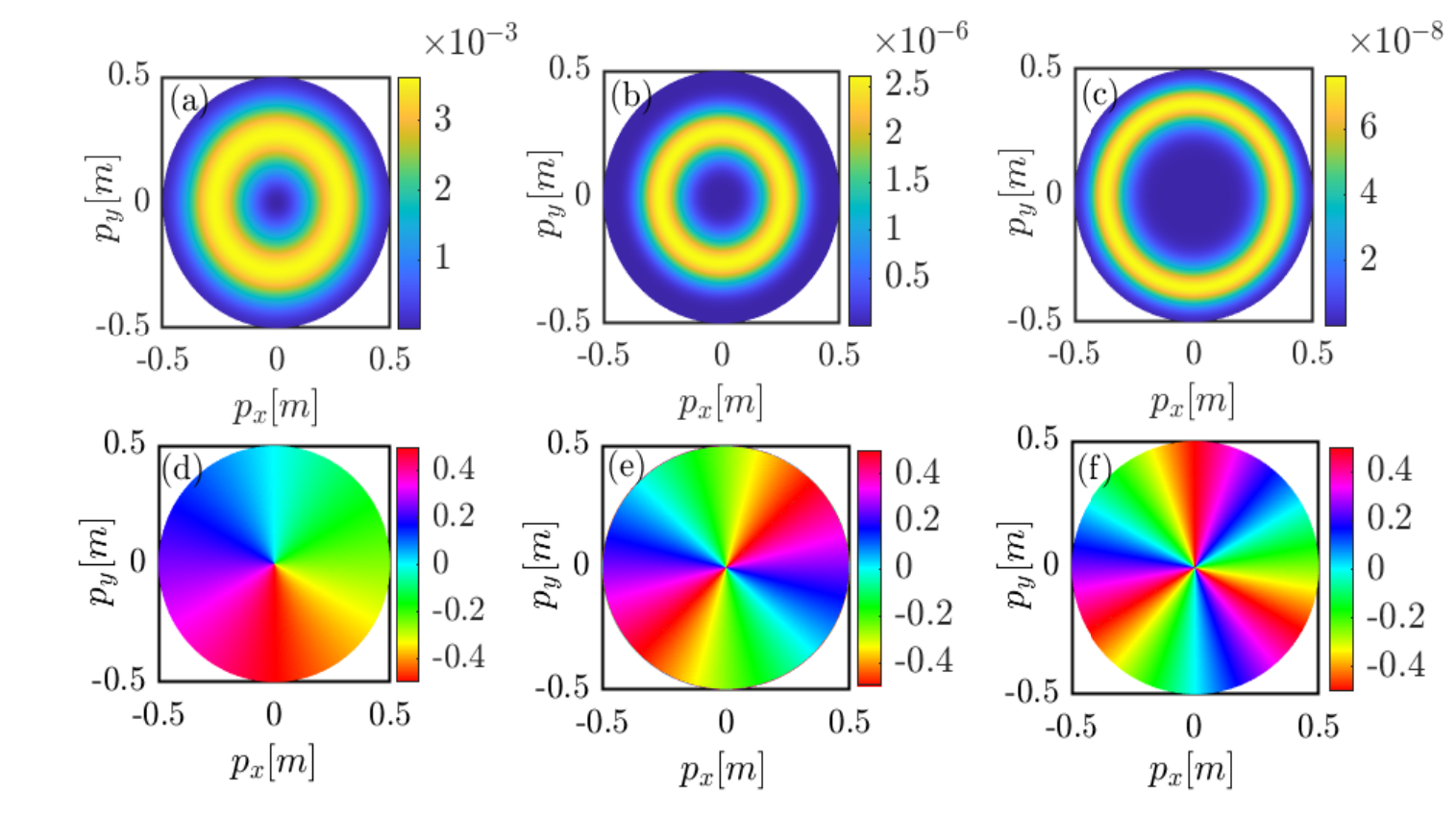}
	\begin{picture}(100,10)		
	\end{picture}
	\caption{Particle distribution function $\vert c_{\beta,\boldsymbol{p}} \vert ^2$ labeled by (a), (b), and (c) as well as the phase $\arg \left[c_{\beta, {\boldsymbol{p}}}\right]/\left(2\pi\right)$ labeled by (d), (e), and (f) for $\omega = 2 m, 1m$, and 0.7$m$ for circular polarization with $\delta = 1$. From left to right are corresponding to one-, two-, and three-photon absorption.}
	\label{Fig:1}
\end{figure*}

In the following of this part, let us review the vortex structures of a complex function.
In quantum mechanics,  probability fluid inherently possess vortices~\cite{BialynickiBirula1992TheoryOQ,Bialynicki-Birula:2016unl}.
Vortex lines are closed loops or continuous curves in three-dimensional momentum space~\cite{Majczak:2022xlv,Hebenstreit:2016xhn,Bliokh:2007ec,Bliokh:2011fi,Bliokh:2012az}. While the complex probability amplitudes $c_{\beta, {\boldsymbol{p}}}$ vanish along these lines,
the phase of the amplitudes around them changes from 0 to 2$\pi l$, where the integer $l$ = ±1, ±2, ... is the topological charge.
In contrast, the nodes form surfaces with vanishing probability, where the phase of $c_{\beta, {\boldsymbol{p}}}$ jumps $\pm \pi$~\cite{Geng2020VortexSI}.
Note that in the two-dimensional plane, vortex lines are usually visualized as individual points, while nodal surfaces behave as curves with zero probability~\cite{Majczak:2022xlv,Majczak:2024hmt,Cajiao:2020}.

We define the gauge-invariant Berry connection of the complex function $c_{\beta, {\boldsymbol{p}}}$ in momentum space~\cite{Majczak:2022xlv,Geng2020VortexSI,Bechler:2023kjx}
\begin{align}\label{Berry}
	{\mathcal{A}} = \frac{\text{Re} \left(c_{\beta, {\boldsymbol{p}}}^*\left[-i\nabla_{\boldsymbol{p}}\right]c_{\beta,
			{\boldsymbol{p}}}\right)}{\lvert c_{\beta, {\boldsymbol{p}}}\rvert^2} = \nabla_{\boldsymbol{p}}\left(\arg \left[c_{\beta, {\boldsymbol{p}}}\right]\right),
\end{align}
where $\nabla_{\boldsymbol{p}}$ represents the gradient with respect to the momentum ${\boldsymbol{p}}$.
The integral along an arbitrary closed curve $\mathcal{C}$ around a singularity of the phase satisfies the following quantized condition
\begin{align}\label{curvature}
	\int_\mathcal{C} {\mathcal{A}}\cdot d\boldsymbol{p} = 2\pi l.
\end{align}

In optics, the topological charge $l$ means of intrinsic orbital angular momentum (IOAM)~\cite{Lloyd:2017,Bliokh:2017uvr}.
The semiclassical analysis for particle packets also shows that the IOAM of a particle is closely related to the Berry connection~\cite{Bliokh:2007ec,Bliokh:2012az}.
In this investigation, taking the primary absorption of photon for example, we find that the number of photons absorbed by the created particles can be presented in the phase. This reveals the angular momentum characteristics of the created particles to some degree.
The momentum spectra and the phase ${\arg}\left[c_{\beta, {\boldsymbol{p}}}\right]/\left(2\pi\right)$ in momentum space are shown in Figs.~\ref{Fig:1}. From Fig.~\ref{Fig:1} we can find that the shape of momentum spectrum as well as the phase are similar to the intensity spectrum and phase of the Laguerre-Gaussian beam~\cite{Allen:1992zz,Zhang:2024ofv}. For the single photon absorption mode [see Figs.~\ref{Fig:1}(a) and (d)], single vortex appears. For two- and three-photon absorption mode [see Figs.~\ref{Fig:1}(b), (c), (e), and (f)], those corresponds to two and three vortices generated, respectively. Most notably, the radius of the absorption ring on the momentum spectrum increases as the number of absorbed photons increases, which is consistent with the phenomenon that large topological charge in the Laguerre-Gaussian mode correspond to large bright ring.

\section{Numerical reasults}\label{Sec:Iv}

In the following numerical calculation, the field amplitude is $E = 0.05E_{\text{cr}}$,
the number of field cycles  $N = 20$, the frequency $\omega = 0.55 m$, and the ellipticity  $\vert \delta \vert = 1$.
These field configurations give the Keldysh parameter of $\gamma = m\omega/eE = 11$, indicating that the system is in the perturbative multiphoton regime~\cite{Keldysh:1965ojf,Blinne:2013via}.
In this regime, the particle-antiparticle pairs not only absorb an unit energy $\omega$ from  each photon  but also the pairs obtain angular momentum by one~\cite{Kohlfurst:2018kxg}.
Interestingly, multiphoton scalar pairs production are constrained by angular momentum conservation, which means that the probability of particle creation vanishes at zero momentum in multiphoton process.

The momentum of the created particles in an oscillating external field can be expressed as ${p}_n = \sqrt{\left({n\omega}/{2}\right)^2 -m^{*2}}$,
where $m^* \approx m \sqrt{1+{1}/{4 \gamma^2}}$ is the effective mass, and $n$ is the number of absorbed photons~\cite{Kohlfurst:2013ura}.
For a clearer presentation of the multiphoton ring structure in the momentum spectra, it is more convenient to present $\lvert c_{\beta, {\boldsymbol{p}}} \rvert$.
The number density of particle can be determined using Eq.~\eqref{distribution}. Both distributions give the same position of $\boldsymbol{p}$ for the $n$-photon absorption.

\subsection{Control of vortex by the helicity}\label{Sec:Iv:A}

\begin{figure}[htbp]
	\includegraphics[scale=0.33]{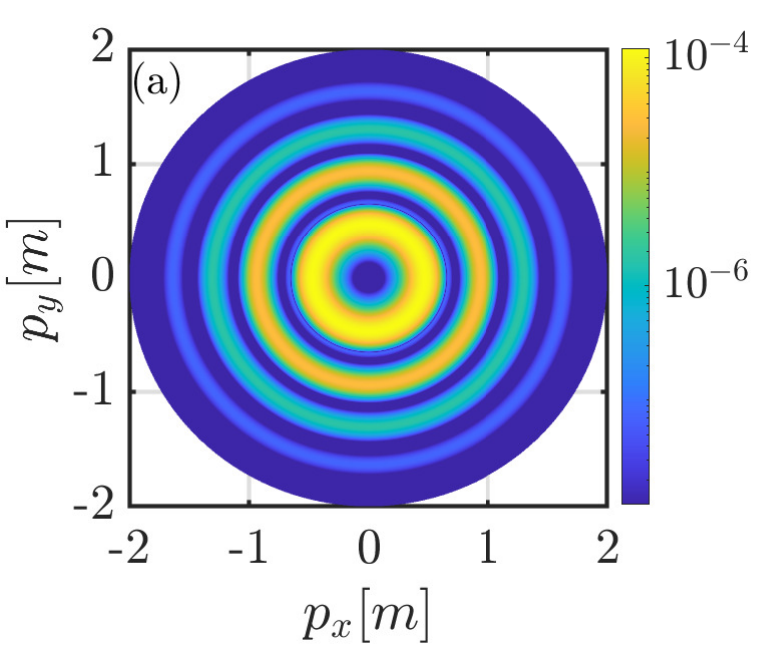}
	\includegraphics[scale=0.33]{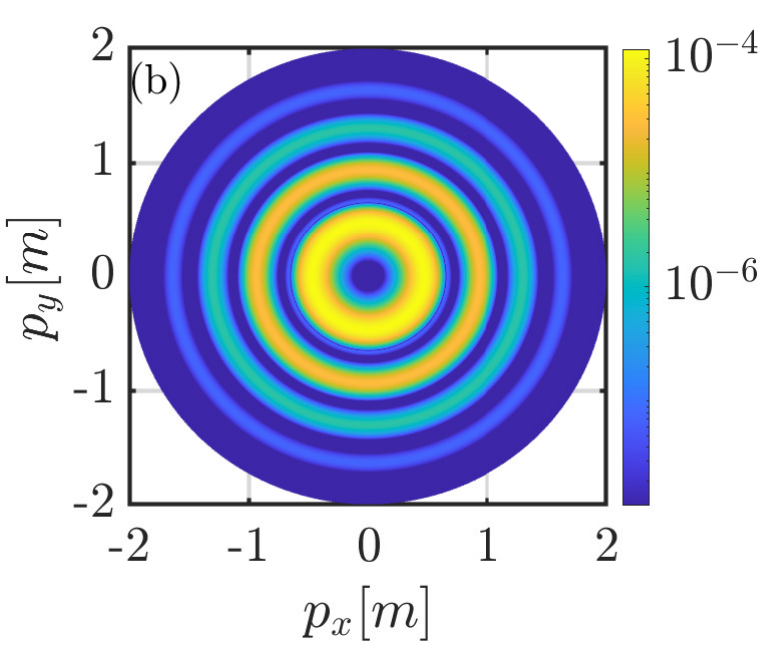}
	\includegraphics[scale=0.33]{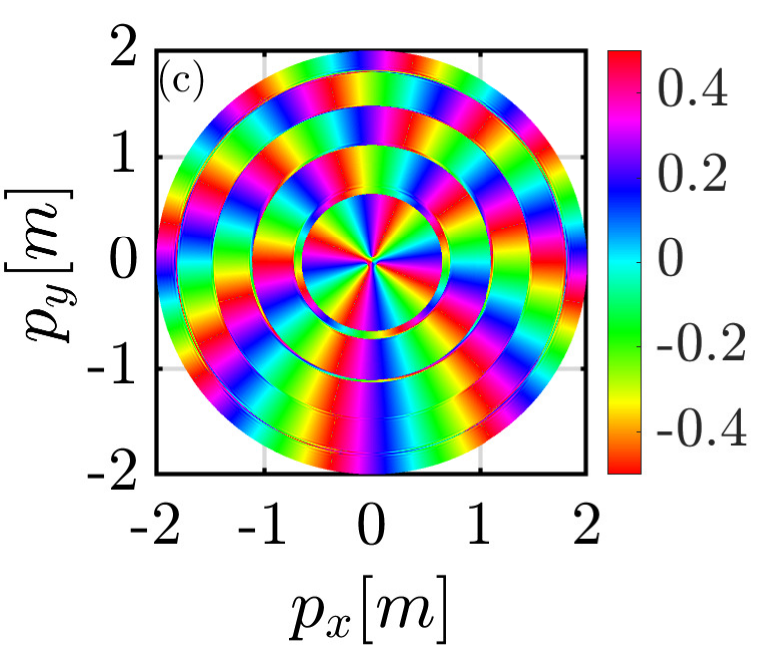}
	\includegraphics[scale=0.33]{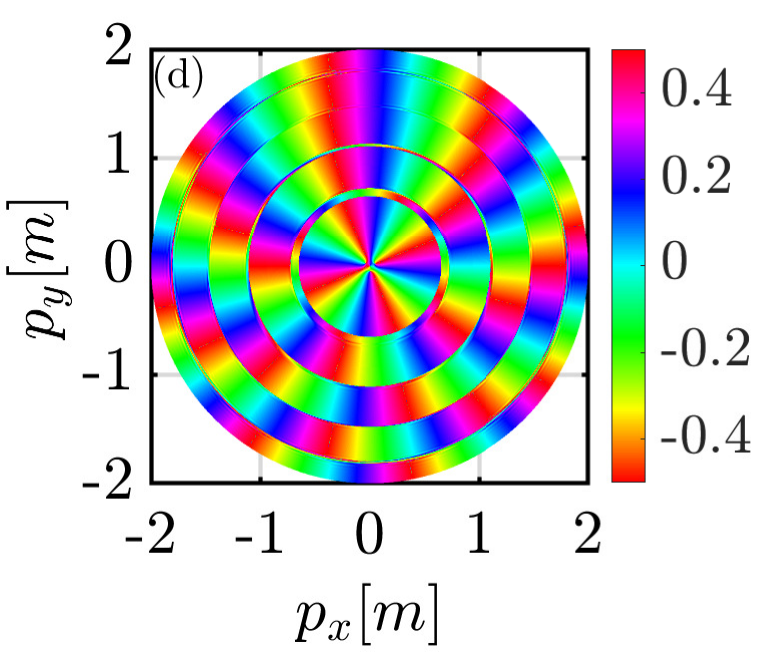}
	\caption{The magnitude of the probability amplitude $\lvert c_{\beta, {\boldsymbol{p}}} \rvert$ for (a) and (b), as well as the phase arg$\left[c_{\beta, {\boldsymbol{p}}}\right]/\left(2\pi\right)$ for (c) and (d) in circularly polarized fields. (a) and (c) correspond to $\delta = 1$ as well as (b) and (d) correspond to $\delta = -1$.}
	\label{Fig:2}
\end{figure}

The magnitude of $\lvert c_{\beta, {\boldsymbol{p}}} \rvert$ and the phase arg$\left[c_{\beta, {\boldsymbol{p}}}\right]/\left(2\pi\right)$ in multiphoton pair production are shown in Fig.~\ref{Fig:2}.
The distributions of the created particles in the polarized plane show observable consistency for $ \delta =  \pm 1$ [see Figs.~\ref{Fig:2}(a) and (b)] in that the left- and right- photon have the same energy for the configuration in Eq.~\eqref{Eq:1}.
Moreover, from Figs.~\ref{Fig:2}(a) and (b), it can be seen that the process corresponding to $n_\text{min}$-photon absorption is often accompanied by $(n_\text{min}+s)$-photon absorption as well, where $n_\text{min} =  \lceil \frac{2m}{\omega} \rceil = 4$ with the ceiling function $\lceil \cdot \rceil$.
The bright rings from inner to outer in Figs.~\ref{Fig:2}(a) and (b) represent four-, five-, six-, and seven-photon absorption which correspond to $s$ = $0$, $1$, $2$, and $3$, respectively.
The similarity in the shape of the momentum spectra to atomic ionization, especially in above-threshold ionization spectra, could be found in Refs.~\cite{Popov:1972,Muller:2009zzf,Popov:2005rp}.

We also find that the vortex direction is counterclockwise in a left-handed field and clockwise in a right-handed field [see Figs.~\ref{Fig:2}(c) and (d)]. This is because the helicity of the photon, as an intrinsic property, can be transferred to the particle's OAM. This OAM is intrinsic, with a status equivalent to the particle's spin.
Quantum mechanics' superposition principle requires that the wave function of the created particles be a superposition of various possible wave packets with angular momentum modes~\cite{Bliokh:2007ec,Bialynicki-Birula:2016unl}. Thus, the wave function of the created particle can be formally written as  $\psi =\sum_{l}\psi_l = \sum_{l} \lvert c_{\beta, {\boldsymbol{p}}}\rvert e^{-il \left(p_r\right)\phi}$ with $p_r = \sqrt{p_x^2 + p_y^2 + p_z^2}$.
Such expansion reveals that $\psi_l$ with nonzero $l$ contain a screw dislocation of the wavefront on the wave packet   $\psi_l \propto e^{-il \left(p_r\right)\phi}$, that is they contain a phase vortex of strength $l$. 

Using the eigenequations for the $l$ model, one can easily obtain $L_z\psi_l = i\frac{\partial}{\partial \phi}\psi_l = l\left(p_r\right)\psi_l$, which has a similar local geometric structure to that of Eq.~\eqref{Berry}. This more clearly shows the consistency between the topological charge and the quantum OAM.
When the particles absorb $n$ left-handed photons, the helicity of the particle is $-l = n$, otherwise it is $+l = n$. Thus, the topological charge serve as a form of IOAM can directly demonstrates that the final-state particle wave packet is born with a distinctly spiral wavefront structure in multiphoton pairs production even in the absence of the spiral structure in the momentum spectra.

More significantly, this property enables the manipulation of the particle IOAM by the external field. The wave of the created particles with different spiral wavefronts can be easily obtained by changing the chirality of the external field.

\subsection{Control of vortex by the absorbed photons} \label{Sec:III:B}

\begin{figure}[htbp]
  \includegraphics[scale=0.68]{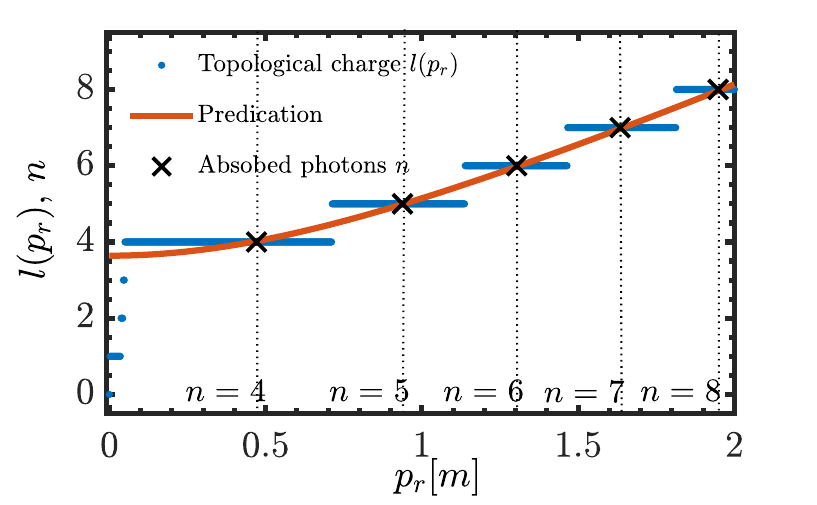}
  \caption{Topological charge (blue staired lines) using Eq.~\eqref{curvature} and the absorbed photons (black crosses) as a function of $p_r$ for $\delta = 1$ at $p_z = 0$. The black vetical dashed lines give the position of kinetic momentum which correspond to the multiphoton peaks. The red curve is the predication of topological charge by using
  energy conservation.}
  \label{Fig:3}
\end{figure}

Based on energy conservation, we can derive an approximate theoretical expression for the topological charge:
$l^\prime \left(p_r\right) = \frac{2}{\omega} \sqrt{m^{*2}+p_r^2}$.
Fig.~\ref{Fig:3} gives the relationship between the topological charge and the number of absorbed photons with the change of the radial momentum $p_r$.
From Fig.~\ref{Fig:3} we find that the approximate theoretical expression, $l^\prime \left(p_r\right)$, gives a better description and the theoretical predication is better in the high energy regime.
Additionally,  the number of absorbed photons and the topological charge are almost perfectly matched.

\begin{figure*}[!t]
	\setlength{\abovecaptionskip}{0.cm}
	\setlength{\belowcaptionskip}{-0.cm}
	 \centering\includegraphics[width=1\linewidth]{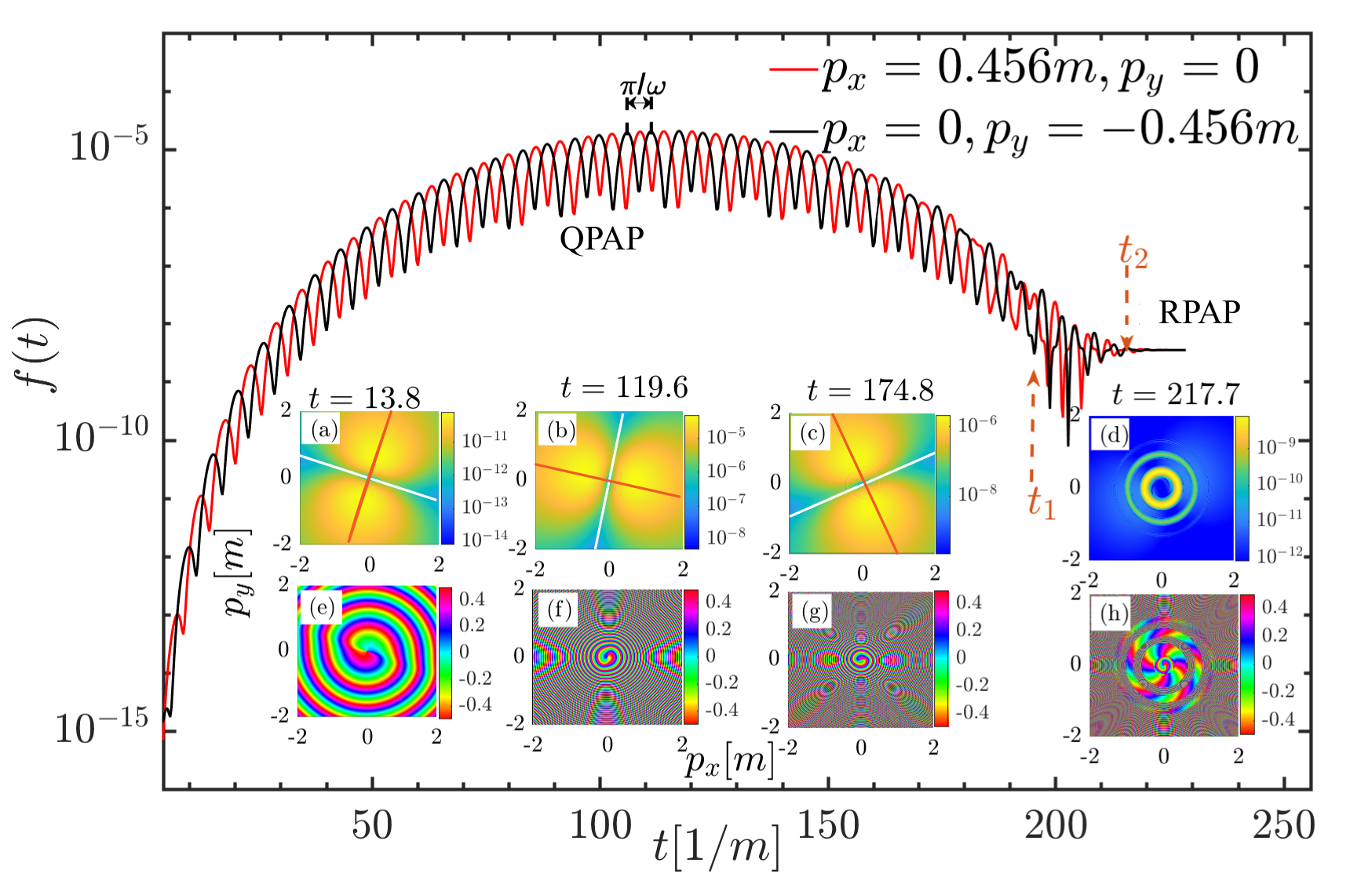}
	\begin{picture}(100,10)
		
	\end{picture}
	\caption{Particle distribution $f(t)$ at different momentum position that correspond to red and black curves as well as the momentum spectra that correspond to (a)-(d) and the phase distribution that correspond to (e)-(h) at the different moment. (a), (b), (c), (e), (f), (g) are QPAP stage. The transient stage corresponds to (d) and (h). The white and orange curves are the lower and higher yield position in (a)-(c). The ellipticity is $\delta = 1$.}
	\label{Fig:4}
\end{figure*}

The relationship between the photon number and the topological charge shows that when the created particles also have  $z$-direction momentum $p_z \neq 0$, leading to the larger $p_r$, the number of absorbed photons may increase, and thus large IOAM particles can be obtained.
For example, when the particles propagate along $z$-direction with $p_z = 0.8m$, the number of absorbed photons increase, leading to an increase in the number of topological charges as well, i.e., $l\to l+1$.
This property gives us two methods of obtaining large IOAM.
Either particles with different topological charges can be obtained at different particles' ejected directions for a constant $p_z$, or particles with large topological charges can be captured in the direction along the pulse propagation when we focus on the strongest $n$-photon ring.

More interestingly, a simple relationship for $s = 0$ between topological charge and external field frequency can also be easily generalized as follows: $l  \propto \omega^{-1}$ or specifically $l = \lceil 2m^*/\omega \rceil$,
which is an inevitable result for the topological charge to be IOAM, see Fig.~\ref{Fig:1}.

\subsection{Time evolution of the vortex}\label{Sec:Iv:C}

For a better understanding of the vortex structures, let us investigate the evolution of the particle and topological charge in the quasiparticle representation.
In external fields, the created particles evolving over time undergo three stages
in the external field as created particle-antiparticle plasma (PAP):
the quasiparticle PAP (QPAP) stage, the transient stage, and the residual PAP (RPAP). 
Note that the geometric phase of wave function of particle acquired in the time evolution plays a key role to lead the nonlinear variation of topological charge in these three stages.

Under the external field Eq.~\eqref{Eq:1}, scalar pairs have the vanishing yield at zero momentum for multiphoton absorption. It reminds us that we can plot the time evolution of the created quasiparticles or/and particle distribution $f(t)$ for choosing two typical momentum cases of $(0.456m,0)$ and $(0,-0.456m)$ on the four-photon ring, which distinguish from $(0,0)$. The results are illustrated in Fig.~\ref{Fig:4}.

From Fig.~\ref{Fig:4}, we observe that the interval between the peaks in the intermediate region of the QPAP stage is half a cycle of the field, i.e., $\pi/\omega$. It is understood as below. Taking $(p_x,p_y) =(0.456m,0)$, for example, the distribution function approximately satisfies $f\left(t\right)\sim \left(\boldsymbol{q}\cdot  \boldsymbol{E}\right)^2 \approx {p_x}^2 E_x^2$ for $p_y = 0$ in QPAP stage due to $E_x \propto \cos\left(\omega t \right), E_y \propto \sin\left(\omega t\right)$ and $A_x \propto \sin\left(\omega t\right), A_y \propto -\cos\left(\omega t\right)$.
The evolution of the envelope increases when time increases, and $E_x^2 \propto \cos^2\left(\omega t\right)$. Therefore, the neighboring peaks in the red curve in Fig.~\ref{Fig:2}, which denotes the created particles distribution $f\left(t\right)$, possess the time interval of $\pi/\omega$.

From Figs.~\ref{Fig:4}(a), (b), and (c), we can see that the momentum spectra of the created particles appears with two peaks similar to the Bessel-Gaussian model~\cite{Saghafi:2001,Tschernig:2024}. The corresponding phase also presents the spiral wave structures [see Figs.~\ref{Fig:4}(e), (f), and (g)].
The above behavior can be understood with the help of the view point of nonlinear dynamical system~\cite{Strogatz:2018}.

In fact we can rewrite Eq.~\eqref{eq:twolevelsys} by considering the complex form of them as $c_{\beta, {\boldsymbol{p}}}\left(t\right) =  C \exp\left(iS_1\right)$ while $c_{\alpha, {\boldsymbol{p}}}\left(t\right) \approx \exp\left(iS_2\right)$ since that $|c_{\alpha, \boldsymbol{p}}(t)|^2-|c_{\beta, \boldsymbol{p}}(t)|^2=1$ and $C\ll 1$.

On one hand, the evolution equation of the momentum distribution is determined by the amplitude $\dot C = -\Omega_{\boldsymbol{p}}\left(t\right) \cos\left(S\right)$, where $S = S_1 + S_2$ and $\Omega_{\boldsymbol{p}}\propto \boldsymbol{p}\cdot \boldsymbol{E}$. It indicates that the extremal minimum can be achieved at either $\boldsymbol{p}\cdot \boldsymbol{E} = 0$ or $\boldsymbol{p} = 0$. The former case, $\boldsymbol{p}\cdot \boldsymbol{E} = 0$, gives the lower yield at $t = 17.4m^{-1}$ along a line with angle $\theta = \arctan\left(p_y/p_x\right) = 165^\circ$ which corresponds to white line in the momentum spectrum, while $\boldsymbol{p} = 0$ shows the result of vanishing pair production at any moment $t$, which is consistent with the requirement of angular momentum conservation. Obviously the other extremal maximum with high pair yield would appear in direction of $\boldsymbol{p} \parallel \boldsymbol{E}$ at $75^\circ$ which corresponds to orange line, see Fig.~\ref{Fig:4}(a). Similarly, the lower yields at $t = 119.6m^{-1}, 174.8m^{-1}$ are $78^\circ$ and $198^\circ$, and the higher yields are $168^\circ$ and $108^\circ$, see the white lines and orange lines in Figs.~\ref{Fig:4}(b) and (c). This is because that the amplitude $C$ is approximately trigonometric function for the fixed $q_r = \sqrt{q_x^2 + q_y^2}$ as the momentum space azimuth $\theta$ changes at the fixed moment $t^*$, there are two extreme values, corresponding to two peaks along the orange curve that is the external field direction. Along the perpendicular direction of electric field, there are two zeros points. When field is either turned off or close to end when $E_x, E_y \sim 0$, then $\Omega_{\boldsymbol{p}} \sim 0$ for all $(p_x,p_y)$, then it produces the different circles for  different fixed $q_r = \sqrt{q_x^2 + q_y^2}$ in the momentum spectrum, which form the shape of rings, see Figs.~\ref{Fig:2}(a) and (b) and also Fig.~\ref{Fig:4}(d).

On the other hand, the phase evolution equation meets that $\dot S = 2\omega_{\boldsymbol{p}} + {\Omega_{\boldsymbol{p}}\sin\left(S\right)}/{C}$. At specific moment $t^*$, we can obtain a constant $\boldsymbol{A}(t^*)$ so that also a constant $\omega_{\boldsymbol{p}}$. Thus for the fixed kinetic momentum $q_r$, the phase $S$ is approximately circular in momentum space as $\theta$ changes when the time approaches the end of the field, where $\Omega_{\boldsymbol{p}} \sim 0$, see Figs.~\ref{Fig:2}(c) and (d) and also Fig.~\ref{Fig:4}(h). In cases of earlier time, however, due to the impact of nonlinear effect of $\Omega_{\boldsymbol{p}}$, it actually produces the spiral line of Archimedes-like for specific $q_r$. For example, the spiral structure is very obvious for the neighbor region of momentum origin $(0,0)$, where $dq_r/d\theta >0$, see Figs.~\ref{Fig:4}(e), (f) and (g). Additionally there are a set of other complex vortex-like substructure that around the central spiral structure, which seems to be an interference pattern between different phases under different momentum.

Interestingly, during the period from $t=174.8m^{-1}$ to $t_1$, the QPAP state becomes unstable and the spiral structure in momentum spectra appears in Fig.~\ref{Fig:4}(c). This can be seen as a modulation of the field by the vacuum fluctuate in the region~\cite{Blaschke:2017igl}.
The tail of the phase spiral becomes elongated and denser due to the quantum mechanical phase continuous accumulation in the real-particles wave function, as shown in Fig.~\ref{Fig:4}(g).

In the transient stage from $t_1$ to $t_2$, the beard shape of the particle spectra is enhanced by the oscillatory effect due to QPAP unstable, as indicated by the orange arrow in the Fig.~\ref{Fig:4}.
The vacuum fluctuations cause the spiral arms in the momentum spectra to elongate and rapidly converge to a ring structure.
Multiple topological charges also manifest in the phase spectra.
The modulation disappears, when the external field weaken.
The transient process fully enters the RPAP region, and the shape of momentum spectra and the vortex structures are the same as in Figs.~\ref{Fig:2}(a) and (c).

\section{Summary}\label{summary}

In this study, we numerically investigate the vortex structure induced by rotating electric fields, as well as the evolution of the vortices and momentum spectra in the quasiparticle representation.

The vortex direction of the final particle is determined by the polarization of the field, suggesting that the topological charge associated with the vortex reflects the IOAM properties of the particle.
The vortex serves as a form of the IOAM of the particle, which is transferred from the photons of the external field. The transfer of angular momentum from $n$ photons to the pairs results in a vortex or topological charge of the particle equal to $n$. The study shows that both the structure and magnitude of the IOAM can be controlled by the external field.
In the quasiparticle representation, we also study the time evolution of the vortex. It is found that the vortex structure remains stable during the QPAP stage. In the transient stage, the vacuum rise and fall leads to the instability of the vortex. Once entering the RPAP region, the momentum spectra of the particle and the vortex structure will rapidly converge to the created particle of the final states.

This work suggests that the transfer of angular momentum in spinless scalar particles during the multiphoton process can be extend to the angular momentum transfer in multiphoton electron-positron pair production with spin. In spinor QED, however, angular momentum conservation may cause the number of vortices in spin-parallel alignment particles to be different from that in spin-antiparallel alignment. Furthermore, the present study also indicates that the topology charge associated to the vortex as a new freedom provides a potential opportunity to detect or/and control the pair production behaviors.

\begin{acknowledgments}
We are grateful to O. Amat and ZL Li for helpful discussions. This work was supported by the National
Natural Science Foundation of China (NSFC) under Grants No. 12375240, No. 11935008, and No. 12104428. The author acknowledges support from the Shandong Provincial Natural Science Foundation, Grants No. ZR2021QA088. The computation was carried outat the HSCC of the Beijing Normal University.	
\end{acknowledgments}


\begin{thebibliography}{99}\suppressfloats

\bibitem{Sauter:1931zz}
F.~Sauter,
Uber das Verhalten eines Elektrons im homogenen elektrischen Feld nach der relativistischen Theorie Diracs,
Z. Phys.~\textbf{69}, 742 (1931).

\bibitem{Heisenberg:1936nmg}
W.~Heisenberg and H.~Euler,
Folgerungen aus der Diracschen Theorie des Positrons,
Z. Phys.~\textbf{98}, 714 (1936).

\bibitem{Schwinger:1951nm}
J.~S.~Schwinger,
On gauge invariance and vacuum polarization,
Phys. Rev.~\textbf{82}, 664 (1951).

\bibitem{PhysRevD.44.1825}
I.~Bialynicki-Birula, P.~Gornicki, and J.~Rafelski,
Phase-space structure of the Dirac vacuum,
Phys. Rev. D \textbf{44}, 1825 (1991).

\bibitem{Fradkin:1991}
E. S. Fradkin, D. M. Gitman and S. M. Shvartsman,
$Quantum$ $Electrodynamics$ $with$ $Unstable$ $Vacuum$ (Springer Verlag, Berlin, 1991).

\bibitem{Ruf:2008ahs}
M.~Ruf, G.~R.~Mocken, C.~M{\"u}ller, K.~Z.~Hatsagortsyan and C.~H.~Keitel,
Pair production in laser fields oscillating in space and time,
Phys. Rev. Lett. \textbf{102}, 080402 (2009).

\bibitem{DiPiazza:2011tq}
A.~Di Piazza, C.~M\"uller, K.~Z.~Hatsagortsyan and C.~H.~Keitel,
Extremely high-intensity laser interactions with fundamental quantum systems,
Rev. Mod. Phys. \textbf{84}, 1177 (2012).

\bibitem{Xie:2017xoj}
B.~S.~Xie, Z.~L.~Li and S.~Tang,
Electron-positron pair production in ultrastrong laser fields,
Matter Radiat. Extremes \textbf{2},~225~(2017).

\bibitem{Keldysh:1965ojf}
L.~V.~Keldysh,
Ionization in the field of a strong electromagnetic wave,
Sov. Phys. JETP 20, 1307 (1965).

\bibitem{Troup:1972xkz}
E. Brezin and C. Itzykson,
Pair production in a vacuum by an alternating field,
Phys. Rev. D \textbf{2}, 1191 (1970).

\bibitem{NBDelone:1994}
N. B. Delone and V.~P.~Krainov,
$Multiphoton$ $Processes$ $in$ $Atoms$ (Springer, Berlin, 1994).

\bibitem{RKopold:2002}
R. Kopold, W. Becker, M. Kleber and G. G. Paulus,
Channel-closing effects in high-order above-threshold ionization and high-order harmonic generation,
J. Phys. B \textbf{35}, 217 (2002).

\bibitem{Ma:2013}
K.~L.~Reid, Annu,
Photoelectron angular distributions,
Rev. Phys. Chem. \textbf{54}, 397 (2003).

\bibitem{BECKER200235}
W. Becker, F. Grasbon, R. Kopold, D.B. Milo{\v s}evi{\'c}, G.G. Paulus and H. Walther,
Above-threshold ionization: from classical features to quantum effects,
Adv. At. Mol. Opt. Phys. \textbf{48} (2002) 35.

\bibitem{Michael:2005}
M. Klaiber, K. Z. Hatsagortsyan and C. H. Keitel,
Above-threshold ionization beyond the dipole approximation,
Phys. Rev. A \textbf{71}, 033408 (2005).

\bibitem{Muller:2009zzf}
C.~M\"uller, K.~Z.~Hatsagortsyan, M.~Ruf, S.~J.~M\"uller, H.~G.~Hetzheim, M.~C.~Kohler and C.~H.~Keitel,
Relativistic nonperturbative above-threshold phenomena in strong laser fields,
Laser Phys. \textbf{19}, 1743 (2009).

\bibitem{Aleksandrov:2021ylw}
I.~A.~Aleksandrov, A.~D.~Panferov and S.~A.~Smolyansky,
Radiation signal accompanying the Schwinger effect,
Phys. Rev. A \textbf{103},~053107 (2021).

\bibitem{Klaiber:023}
A.~Eckey, M.~Klaiber, A.~B.~Voitkiv, and C. M{\"u}ller,
Relativistic strong-field ionization of hydrogenlike atomic systems in constant crossed electromagnetic fields, Phys. Rev. A \textbf{107}, 033113 (2023).

\bibitem{Klaiber:2024}
M.~Klaiber, K.~Z.~Hatsagortsyan, C.~H.~Keitel,
Atomic polarization and Stark-shift in relativistic strong field ionization, [arXiv.2403.11285 [{physics.atom-ph}]].

\bibitem{Kohlfurst:2018kxg}
C.~Kohlf\"urst,
Spin states in multiphoton pair production for circularly polarized light,
Phys. Rev. D \textbf{99},~096017 (2019).

\bibitem{Blinne:2013via}
A.~Blinne and H.~Gies,
Pair production in rotating electric fields,
Phys. Rev. D \textbf{89}, 085001 (2014).

\bibitem{Li_2015}
Z.~L.~Li, D.~Lu, B.~S.~Xie, B.~F.~Shen, L.~B.~Fu and J.~Liu,
Nonperturbative signatures in pair production for general elliptic polarized fields,
Europhys. Lett.~\textbf{110}, 51001 (2015).

\bibitem{Fillion-Gourdeau:2017uss}
F.~Fillion-Gourdeau, F.~Hebenstreit, D.~Gagnon and S.~MacLean,
Pulse shape optimization for electron-positron production in rotating fields,
Phys. Rev. D \textbf{96}, 016012 (2017).

\bibitem{Geng2020VortexSI}
Lei Geng, F. Cajiao V{\'e}lez, J. Z. Kami{\'n}ski, Liang-You Peng and Katarzyna Krajewska,
Vortex structures in photodetachment by few-cycle circularly polarized pulses,
Phys. Rev. A \textbf{102}, 043117 (2020).

\bibitem{Ngoko:2015}
J. M. Ngoko Djiokap, S.~X.~Hu,~L.~B.~Madsen,~N.~L.~Manakov,~A.~V.~Meremianin~and~Anthony~F.~Starace,
Electron Vortices in Photoionization by Circularly Polarized Attosecond Pulses,
Phys. Rev. Lett. \textbf{115}, 113004 (2015).

\bibitem{Pengel:2017}
D. Pengel, S. Kerbstadt, D. Johannmeyer, L. Englert, T. Bayer and M. Wollenhaupt,
Electron vortices in femtosecond multiphoton ionization,
Phys. Rev. Lett. \textbf{118}, 053003 (2017).

\bibitem{Li:2017qwd}
Z.~L.~Li, Y.~J.~Li and B.~S.~Xie,
Momentum vortices on pairs production by two counter-rotating fields,
Phys. Rev. D \textbf{96}, 076010 (2017).

\bibitem{Li:2018hzi}
Z.~L.~Li, B.~S.~Xie and Y.~J.~Li,
Vortices in multiphoton pair production by two-color rotating laser fields,
J. Phys. B \textbf{52}, 025601 (2018).

\bibitem{Li:2019rex}
Z.~L.~Li, B.~S.~Xie and Y.~J.~Li,
Boson pair production in arbitrarily polarized electric fields,
Phys. Rev. D \textbf{100}, 076018 (2019).

\bibitem{Majczak:2022xlv}
M.~M.~Majczak, F.~Cajiao V\'elez, J.~Z.~Kami\'nski and K.~Krajewska,
Carrier-envelope-phase and helicity control of electron vortices and spirals in photodetachment,
Opt. Express \textbf{30}, 43330-43341 (2022).

\bibitem{Hu:2023pmz}
L.~N.~Hu, O.~Amat, L.~Wang, A.~Sawut, H.~H.~Fan and B.~S.~Xie,
Momentum spirals in multiphoton pair production revisited,
Phys. Rev. D \textbf{107}, 116010 (2023).

\bibitem{Yusoff:2024}
M.~A.~H.~B.~Md~Yusoff and J. M. Ngoko Djiokap,
Time-delay control of reversible electron spirals using arbitrarily chirped attosecond pulses,
Phys. Rev. A, \textbf{109}, 023107, (2024).

\bibitem{PMDirac:1931}
P.~A.~M.~Dirac,
Quantised singularities in the electromagnetic field
Proc. R. Soc. A \textbf{133}, 60 (1931).

\bibitem{Beth:1936zz}
R.~A.~Beth,
Mechanical detection and measurement of the angular momentum of light,
Phys. Rev. \textbf{50}, 115-125 (1936).

\bibitem{Aharonov}
Y.~Aharonov~and~D.~Bohm,
Significance of electromagnetic potentials in the quantum theory,
Phys. Rev. \textbf{115}, 485 (1959); M.~Peshkin~and~A.~Tonomura, $The$ $Aharonov$-$Bohm$ $Effect$
(Springer, Berlin, 1989).

\bibitem{berry}
M.~V. Berry,
Quantal phase factors accompanying adiabatic changes,
Proc. R. Soc. A \textbf{392}, 45 (1984).

\bibitem{Hebenstreit:2016xhn}
F.~Hebenstreit,
Vortex formation and dynamics in two-dimensional driven-dissipative condensates,
Phys. Rev. A \textbf{94}, 063617 (2016).

\bibitem{Bialynicki-Birula:2016unl}
I.~Bialynicki-Birula and Z.~Bialynicka-Birula,
Relativistic electron wave packets carrying angular momentum,
Phys. Rev. Lett. \textbf{118}, 114801 (2017).

\bibitem{Allen:1992zz}
L.~Allen, M.~W.~Beijersbergen, R.~J.~C.~Spreeuw and J.~P.~Woerdman,
Orbital angular momentum of light and the transformation of Laguerre-Gaussian laser modes,
Phys. Rev. A \textbf{45}, 8185 (1992).

\bibitem{BialynickiBirula1992TheoryOQ}
I.~Bialynicki-Birula, M.~Cieplak and J.~Kami\'nski,
$Theory$ $of$ $Quanta$ (Oxford University Press, New York, 1992).

\bibitem{Bechler:2023kjx}
A.~Bechler, F.~Cajiao V\'elez, K.~Krajewska and J.~Z.~Kami\'nski,
Vortex structures and momentum sharing in dynamic Sauter-Schwinger process,
Acta Phys. Polon. A \textbf{143}, S18 (2023).

\bibitem{Majczak:2024hmt}
M.~M.~Majczak, K.~Krajewska, J.~Z.~Kami\'nski and A.~Bechler,
Scattering matrix approach to dynamical Sauter-Schwinger process:~spin- and helicity-resolved momentum distributions,
[arXiv:2403.15206 [quant-ph]].

\bibitem{Fan:2024vqi}
H.~H.~Fan, L.~N.~Hu, S.~Tang, Z.~L.~Li and B.~S.~Xie,
Application of partial wave analysis in multiphoton pair production,
Phys. Rev. D \textbf{110}, 076028 (2024).

\bibitem{Lloyd:2017}
S.~M.~Lloyd,~M.~Babiker,~G.~Thirunavukkarasu~and~J.~Yuan,
Electron vortices:~beams with orbital angular momentum,
Rev. Mod. Phys. \textbf{89}, 035004 (2017).

\bibitem{Bliokh:2017uvr}
K.~Y.~Bliokh, I.~P.~Ivanov, G.~Guzzinati, L.~Clark, R.~Van Boxem, A.~B\'ech\'e, R.~Juchtmans, M.~A.~Alonso, P.~Schattschneider and F.~Nori, \textit{et al.}
Theory and applications of free-electron vortex states,
Phys. Rept. \textbf{690}, 1 (2017).

\bibitem{Bliokh:2007ec}
K.~Y.~Bliokh, S.~Savel\'ev and F.~Nori,
Semiclassical dynamics of electron wave packet states with phase vortices,
Phys. Rev. Lett. \textbf{99}, 190404 (2007).

\bibitem{Kluger:1998bm}
Y.~Kluger, E.~Mottola and J.~M.~Eisenberg,
The quantum Vlasov equation and its Markov limit,
Phys. Rev. D \textbf{58}, 125015 (1998).

\bibitem{Dumlu:2011rr}
C.~K.~Dumlu and G.~V.~Dunne,
Interference effects in Schwinger cacuum pair production for time-dependent laser pulses,
Phys. Rev. D \textbf{83}, 065028 (2011).

\bibitem{Akkermans:2011yn}
E.~Akkermans and G.~V.~Dunne,
Ramsey fringes and time-domain multiple-slit interference from vacuum,
Phys. Rev. Lett. \textbf{108}, 030401 (2012).

\bibitem{Dunne:2022zlx}
G.~V.~Dunne, A.~Florio and D.~E.~Kharzeev,
Entropy suppression through quantum interference in electric pulses,
Phys. Rev. D \textbf{108}, L031901 (2023).

\bibitem{Blaschke:2014fca}
D.~Blaschke, L.~Juchnowski, A.~Panferov and S.~Smolyansky,
Dynamical Schwinger effect: properties of the \ensuremath{e^+}\ensuremath{e^-} plasma created from vacuum in strong laser fields,
Phys. Part. Nucl. \textbf{46}, 797 (2015).

\bibitem{Blaschke:2017igl}
D.~B.~Blaschke, S.~A.~Smolyansky, A.~Panferov and L.~Juchnowski,
Particle production in strong time-dependent fields,
[arXiv:1704.04147 [hep-ph]].

\bibitem{Bliokh:2011fi}
K.~Y.~Bliokh, M.~R.~Dennis and F.~Nori,
Relativistic electron vortex beams: angular momentum and spin-orbit interaction,
Phys. Rev. Lett. \textbf{107}, 174802 (2011).

\bibitem{Bliokh:2012az}
K.~Y.~Bliokh and F.~Nori,
Spatio-temporal vortex beams and angular momentum,
Phys. Rev. A \textbf{86}, 033824 (2012).

\bibitem{Cajiao:2020}
F. Cajiao V{\'e}lez, Lei Geng, J. Z. Kami{\'n}ski, Liang-You Peng and K. Krajewska
Vortex streets and honeycomb structures in photodetachment driven by linearly polarized few-cycle laser pulses, Phys. Rev. A \textbf{102}, 043102 (2020).

\bibitem{Zhang:2024ofv}
De-Sheng Zhang, Xue-Ren Hong, Xiao-Bo Zhang, Rong-An Tang and Bai-Song Xie,
Generation of the vortex terahertz radiation by the interaction of two-color Laguerre-Gaussian laser with plasmas in the presence of a static magnetic fieldPhys,
Phys. Plasmas \textbf{31}, 073106 (2024).

\bibitem{Kohlfurst:2013ura}
C.~Kohlf\"urst, H.~Gies and R.~Alkofer,
Effective mass signatures in multiphoton pair production,
Phys. Rev. Lett.~\textbf{112},~050402 (2014).

\bibitem{Popov:1972}
V. S. Popov,
Imaginary time method in atom ionization and pair production problems,
Zh. Eksp. Teor. Fiz. \textbf{63}, 1586, 1972.

\bibitem{Popov:2005rp}
V.~S.~Popov,
Imaginary-time method in quantum mechanics and field theory,
Phys. Atom. Nucl. \textbf{68}, 686 (2005).

\bibitem{Saghafi:2001}
S. Saghafi, C.~J.~R.~Sheppard and~J.~A.~Piper,
Characterising elegant and standard Hermite–Gaussian beam modes,
Opt. Commun.~\textbf{191}, 173 (2001).

\bibitem{Tschernig:2024}
K. Tschernig, D. Guacaneme, O. Mhibik,  I. Divliansky and M. A. Bandres,
Observation of Boyer-Wolf Gaussian modes, Nat. Commun. \textbf{15}, 5301 (2024).

\bibitem{Strogatz:2018}
S.~H.~Strogatz,
$Nonlinear$ $Dynamics$ $and$ $Chaos$: $With$ $Applications$ $to$ $Physics$, $Biology$, $Chemistry$, $and$ $Engineering$,
(CRC Press, 2018) (2nd~ ed.).

\bibitem{Strobel:2013vza}
E.~Strobel and S.~S.~Xue,
Semiclassical pair production rate for time-dependent electrical fields with more than one component: WKB-approach and world-line instantons,
Nucl. Phys. B \textbf{886}, 1153 (2014).

\end{thebibliography}
\end{document}